\documentclass[letter]{jpsj2} 
\title{Fermi surface in BaNi$_2$P$_2$}

\author{Taichi \textsc{Terashima}$^{1, 4}$, Motoi \textsc{Kimata}$^{1}$, Hidetaka \textsc{Satsukawa}$^{1}$, Atsushi \textsc{Harada}$^{1}$, Kaori \textsc{Hazama}$^{1}$, Motoharu \textsc{Imai}$^{1, 4}$, Shinya \textsc{Uji}$^{1, 4}$, Hijiri \textsc{Kito}$^{2, 4}$, Akira \textsc{Iyo}$^{2, 4}$, Hiroshi \textsc{Eisaki}$^{2, 4}$, and Hisatomo \textsc{Harima}$^{3, 4}$}

\inst{$^{1}$National Institute for Materials Science, Tsukuba, Ibaraki 305-0003\\
$^{2}$National Institute of Advanced Industrial Science and Technology (AIST), Tsukuba, Ibaraki 305-8568\\
$^{3}$Department of Physics, Graduate School of Science, Kobe University, Kobe, Hyogo 657-8501\\
$^{4}$JST, Transformative Research-Project on Iron Pnictides (TRIP), Chiyoda, Tokyo 102-0075}

\abst{We report measurements of the de Haas-van Alphen (dHvA) oscillation and a band structure calculation for the pnictide superconductor BaNi$_2$P$_2$, which is isostructural to BaFe$_2$As$_2$, the mother compound of the iron-pnictide high-$T_c$ superconductor (Ba$_{1-x}$K$_x$)Fe$_2$As$_2$.  Six dHvA-frequency branches with frequencies up to $\sim$8 kT were observed, and they are in excellent agreement with results of the band-structure calculation.  The determined Fermi surface is large, enclosing about one electron and hole per formula unit, and three-dimensional.  This is in contrast to the small two-dimensional Fermi surface expected for the iron-pnictide high-$T_c$ superconductors.  The mass enhancement is about two.}

\kword{Fermi surface, BaNi$_2$P$_2$, de Haas-van Alphen effect, electronic structure, pnictide superconductor}

\begin{document}
\maketitle

\def\degree{\kern-.2em\r{}\kern-.3em}

The discoveries of superconductivity in LaFePO ($T_c \sim 4$ K),\cite{Kamihara06JACS} LaNiPO ($T_c \sim 3$ K),\cite{Watanabe07IC} and LaFeAsO$_{1-x}$F$_x$ ($T_c = 26$ K)\cite{Kamihara08JACS} by Kamihara, Watanabe, Hosono, and coworkers have initiated enthusiastic search for superconductivity in layered iron and nickel pnictides.  The superconducting transition temperature $T_c$ has quickly been raised to 43 K by the application of pressure\cite{Takahashi08Nature} and to 54--56 K by the use of high-pressure synthesis and/or rare-earth/actinoid substitution.\cite{Kito08JPSJ, Ren08CPL, Yang08SST, Wang08EPL}  Different crystal structures have also been found to be compatible with this new type of high-temperature superconductivity: (Ba$_{1-x}$K$_x$)Fe$_2$As$_2$ ($T_c = 38$ K), \cite{Rotter08PRL} $\alpha$-FeSe ($T_c = 8$ K),\cite{Hsu08PNAS} and LiFeAs ($T_c = 18$ K),\cite{Wang08SSC} for example, although $\alpha$-FeSe is not a pnictide.  The key ingredient of the high-temperature superconductivity seems to be the two-dimensional square lattice of Fe$^{2+}$ ions.  Since iron and nickel are archetypal magnetic elements and parent compounds, such as LaFeAsO and BaFe$_2$As$_2$, exhibit magnetic transitions, it is suggested that the paring mechanism is of magnetic origin.

It is interesting to determine why the combination of Fe and As is unique.  Among the four possible combinations, i.e.,  (Fe, Ni)  $\times$ (P, As), only the FeAs combination gives high $T_c$'s.  To address this issue, we have initiated a fermiology study of iron and nickel pnictides, and here, we report the first results, namely, the results of de Haas-vam Alphen (dHvA) measurements and a band-structure calculation performed for BaNi$_2$P$_2$.  BaNi$_2$P$_2$ crystallizes in the ThCr$_2$Si$_2$ structure as BaFe$_2$As$_2$ and becomes superconducting below $T_c \sim 3$ K.\cite{Keimes97ZAAC, Mine08SSC}

BaNi$_2$P$_2$ single crystals were prepared by high-pressure synthesis from the constituent elements: the synthesis temperature and pressure were 1250\degree C and 1 GPa, respectively.  The typical in-plane residual resistivity, residual resistivity ratio (RRR), and $T_c$ of the crystals have been reported to be 5 $\mu \Omega$cm, 18, and 2.51 K, respectively.\cite{Tomioka08JPSJS}  The dHvA torque measurements were performed on five single-crystal samples (samples \#1, \#2, \#11, WM, and HM) with typical dimensions of 0.4 $\times$ 0.2 $\times$ 0.02 mm$^3$ by using piezoresistive microcantilevers.  The crystal axes of samples \#2 and WM were determined using X-ray diffractometry with a three-circle goniometer combined with a CCD area detector.  For the other samples, the axes were identified by the inspection of the crystal shapes and surfaces: crystals are generally elongated along a $<110>$ axis, and zigzags, each short segment of which is parallel to a $<100>$ axis, run parallel to $<110>$ axes on \{001\} surfaces.  Samples \#1, \#2, and \#11 were measured up to a magnetic field $B$ of 17.8 T in a dilution refrigerator, while sample WM and samples \#2 and HM were measured in a helium-3 refrigerator up to 25 and 36 T provided by the water-cooled resistive and hybrid magnets, respectively, at the Tsukuba Magnet Laboratory of the NIMS.\cite{TML}  The field was rotated in the (010) and $(1\overline10)$ planes, and the field angles $\theta_{(010)}$ and $\theta_{(1\overline10)}$ were measured from the [001] axis. 

\begin{figure}[tb]
\begin{center}
\includegraphics[width=6.5cm]{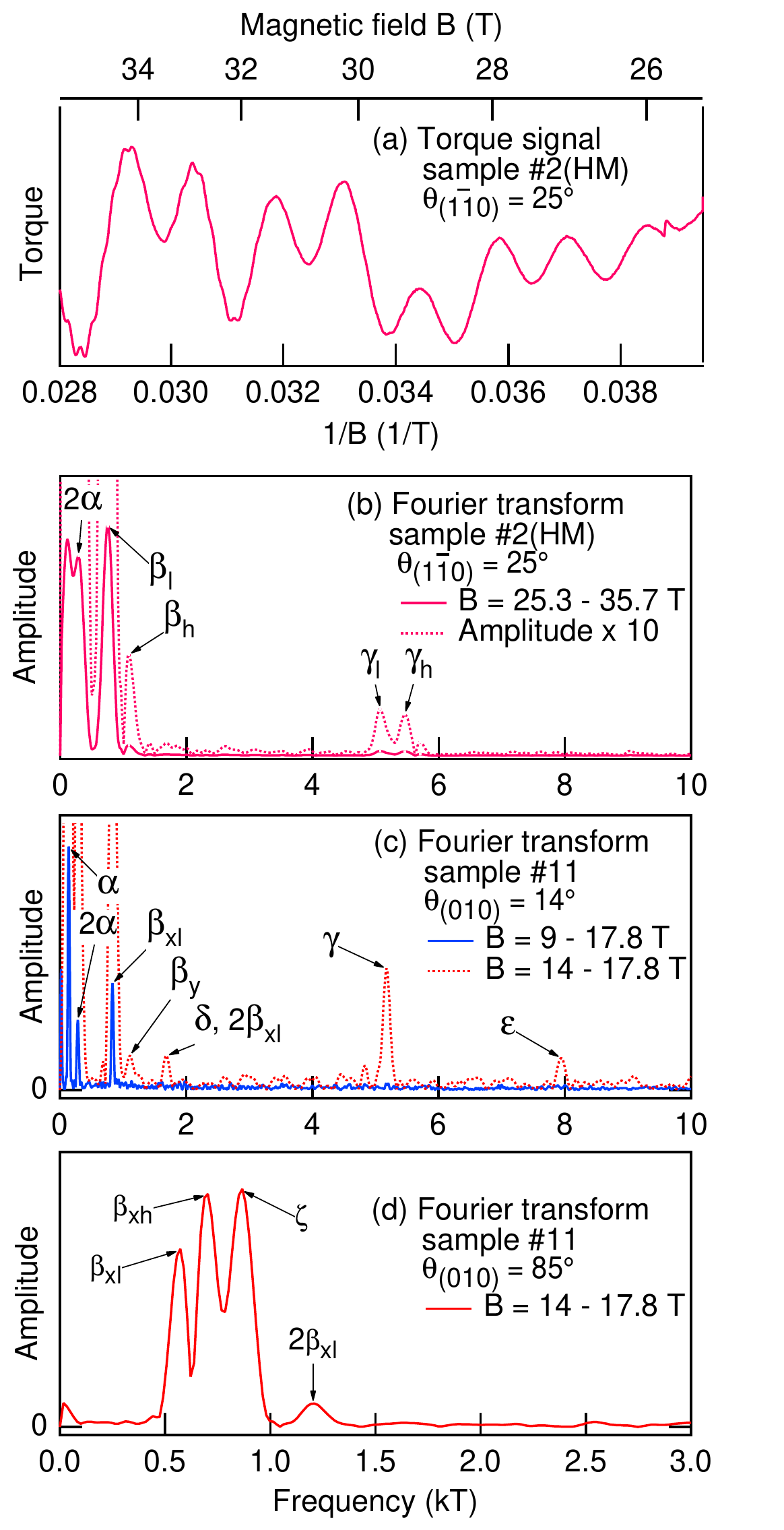}
\end{center}
\caption{(Color online) Examples of (a) the dHvA torque oscillation in BaNi$_2$P$_2$ and (b--d) the Fourier transforms of dHvA oscillations (in inverse field).  In (a), a smooth background has been subtracted.  In (b--d), dHvA frequencies are labeled with Greek letters, and ''$2\alpha$'', etc. indicate the second harmonic of $\alpha$ and so on.  The field directions are indicated in the figures.}
\label{f1}
\end{figure}

\begin{figure}[tb]
\begin{center}
\includegraphics[width=17.5cm]{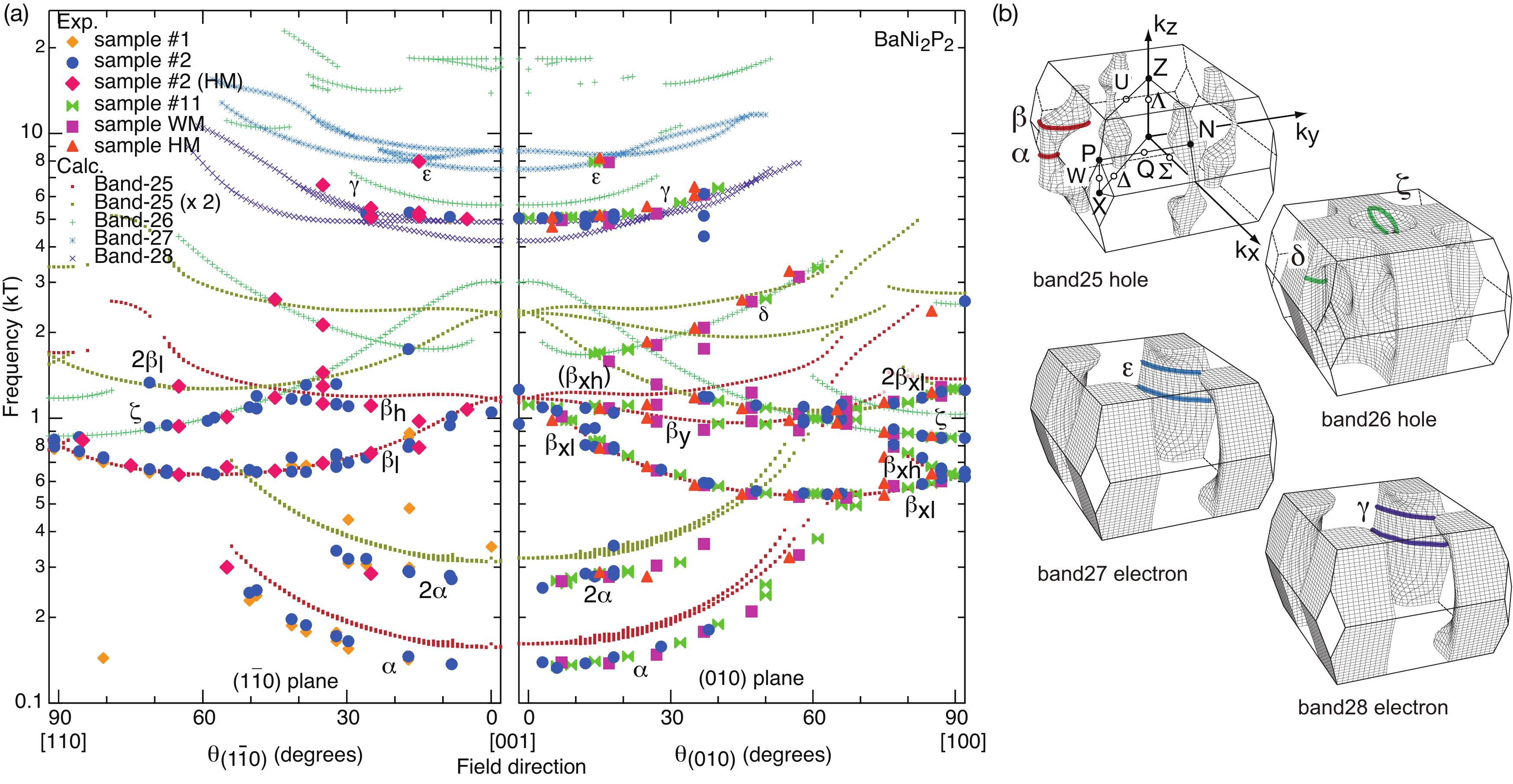}
\end{center}
\caption{(Color online) (a) Angular dependence of the experimental dHvA frequencies compared with the calculated ones.  For the calculated band-25 frequencies, the second harmonics are also shown.  (b) Calculated Fermi surface of BaNi$_2$P$_2$.  Some of the possible dHvA orbits are indicated.  Points of symmetry (solid circles) and lines of symmetry (open circles) in the Brillouin zone are explained in the top left figure.}
\label{f2}
\end{figure}

Figure~\ref{f1}(a) shows an example of the dHvA torque oscillation.  This measurement was performed on sample \#2 in the hybrid magnet up to 36 T.  Fast oscillations with small amplitudes are visible near the high-field side of the data.  Figure~\ref{f1}(b) shows the Fourier transform of the oscillation in an inverse field.  The frequencies $\gamma_l$ and $\gamma_h$ near $F=5$~kT are clearly resolved in addition to three relatively low frequencies, i.e, the second harmonic of $\alpha$ (2$\alpha$), $\beta_l$, and $\beta_h$.  Figures~\ref{f1}(c) and \ref{f1}(d) show two more examples of the Fourier transforms of the dHvA oscillations observed in sample \#11.  Figure~\ref{f2}(a) shows the angular dependence of the dHvA frequencies.  The six fundamental-frequency branches $\alpha$, $\beta$, $\gamma$, $\delta$, $\epsilon$, and $\zeta$ are clearly visible, among which $\beta$ is composed of the symmetry-related sub-branches $\beta_l$, $\beta_h$, $\beta_{xl}$, $\beta_{xh}$, and $\beta_y$, and $\gamma$ appears as a doublet ($\gamma_l$ and $\gamma_h$) for some field directions.  The second harmonic of $\alpha$ is also observed in a wide range of field directions.  The effective mass $m^*$ and Dingle temperature $x_D^*$ associated with dHvA orbits were estimated for the selected field directions from the temperature and field dependences of dHvA oscillation amplitudes, respectively,\cite{Shoenberg84} and are listed in Table~\ref{t1}, where carrier mean free paths $l$ are also derived from $F$, $m^*$, and $x_D^*$.

\begin{table}[tb]
\begin{scriptsize}
\caption{Experimental and calculated Fermi surface parameters of BaNi$_2$P$_2$.}
\label{t1}
\begin{tabular}{cccccccccccc}
\hline
 & \multicolumn{6}{c}{Experiment} & & \multicolumn{3}{c}{Calculation} \\
\cline{2-7} \cline{9-11}
Field direction & Branch & Sample & \multicolumn{1}{c}{$F$ (kT)} & \multicolumn{1}{c}{$m^*/m_e$} & \multicolumn{1}{c}{$x_D^*$ (K)} & \multicolumn{1}{c}{$l$ ($\mu$m)} & & Band & \multicolumn{1}{c}{$F$ (kT)} & \multicolumn{1}{c}{$m_{band}/m_e$} & \multicolumn{1}{c}{$m^*/m_{band}$}\\
\hline
 $\theta_{(1\overline10)} = 32$\degree & $\alpha$ & \#1 & 0.17 & 0.53(5) & \multicolumn{1}{c}{$\sim$6} & \multicolumn{1}{c}{$\sim$0.03} & & 25 & 0.20 & 0.28 & 1.89\\ 
 & & \#2 & 0.17 & 0.54(4) & 5.7(9) & 0.033(3) & & 25 & 0.20 & 0.28 & 1.92 \\[1ex] 
 & $\beta_l$ & \#1 & 0.72 & 0.79(5) & 2.7(9) & 0.10(3) & & 25 & 0.71 & 0.44 & 1.81\\ 
 & & \#2 & 0.68 & 0.75(5) & 6(1) & 0.047(6) & & 25 & 0.71 & 0.44 & 1.72\\ 
\\
 $\theta_{(010)} = 17$\degree & $\alpha$ & WM & 0.14 & 0.41(3) & 6(1) & 0.037(4) & & 25 & 0.17 & 0.25 & 1.66\\ 
 & $\beta_{xl}$ & WM & 0.78 & 0.96(4) & 6.7(6) & 0.033(2) & & 25 & 0.78 & 0.51 & 1.89\\ 
 & $\gamma$ & WM & 5.08 & 1.50(9) & 8(2) & 0.048(6) & & 28 & 5.00 & 0.82 & 1.84\\
 \\ 
 $\theta_{(010)} = 83$\degree & $\beta_{xl}$ & \#2 & 0.59 & 0.66(5) & 7.8(9) & 0.036(2) & & 25 & 0.58 & 0.35 & 1.90\\
 & $\beta_{xh}$ & \#2 & 0.73 & 0.6(1) & & & & 25 & 0.69 & 0.46 & 1.3\\
 & $\zeta$ & \#2 & 0.87 & 0.99(4) & 7.3(5) & 0.032(1) & & 26 & 0.87 & 0.35 & 2.8\\  
\\
 $\theta_{(010)} = 85$\degree & $\beta_{xl}$ & \#11 & 0.59 & 0.65(3) & 9.6(8) & 0.030(1) & & 25 & 0.59 & 0.36 & 1.83\\
 & $\beta_{xh}$ & \#11 & 0.68 & 0.77(3) & 10.7(9) & 0.025(1) & & 25 & 0.66 & 0.43 & 1.80\\
 & $\zeta$ & \#11 & 0.86 & 1.09(4) & 5.6(9) & 0.037(5) & & 26 & 0.87 & 0.35 & 3.1\\
 \hline
\end{tabular}
\end{scriptsize}
\end{table}

The electronic band structure of BaNi$_2$P$_2$ was calculated within the local density approximation (LDA) by using a full potential LAPW (FLAPW) method.  We used the program codes TSPACE\cite{Yanase1995} and KANSAI-06.  The following experimental lattice parameters\cite{Keimes97ZAAC} were used for the calculation: $a$ = 3.947\,\AA, $c$ = 11.820\,\AA, and $z$ = 0.3431 for the 4e sites of P.  

\begin{figure}[tb]
\begin{center}
\includegraphics[width=6.5cm]{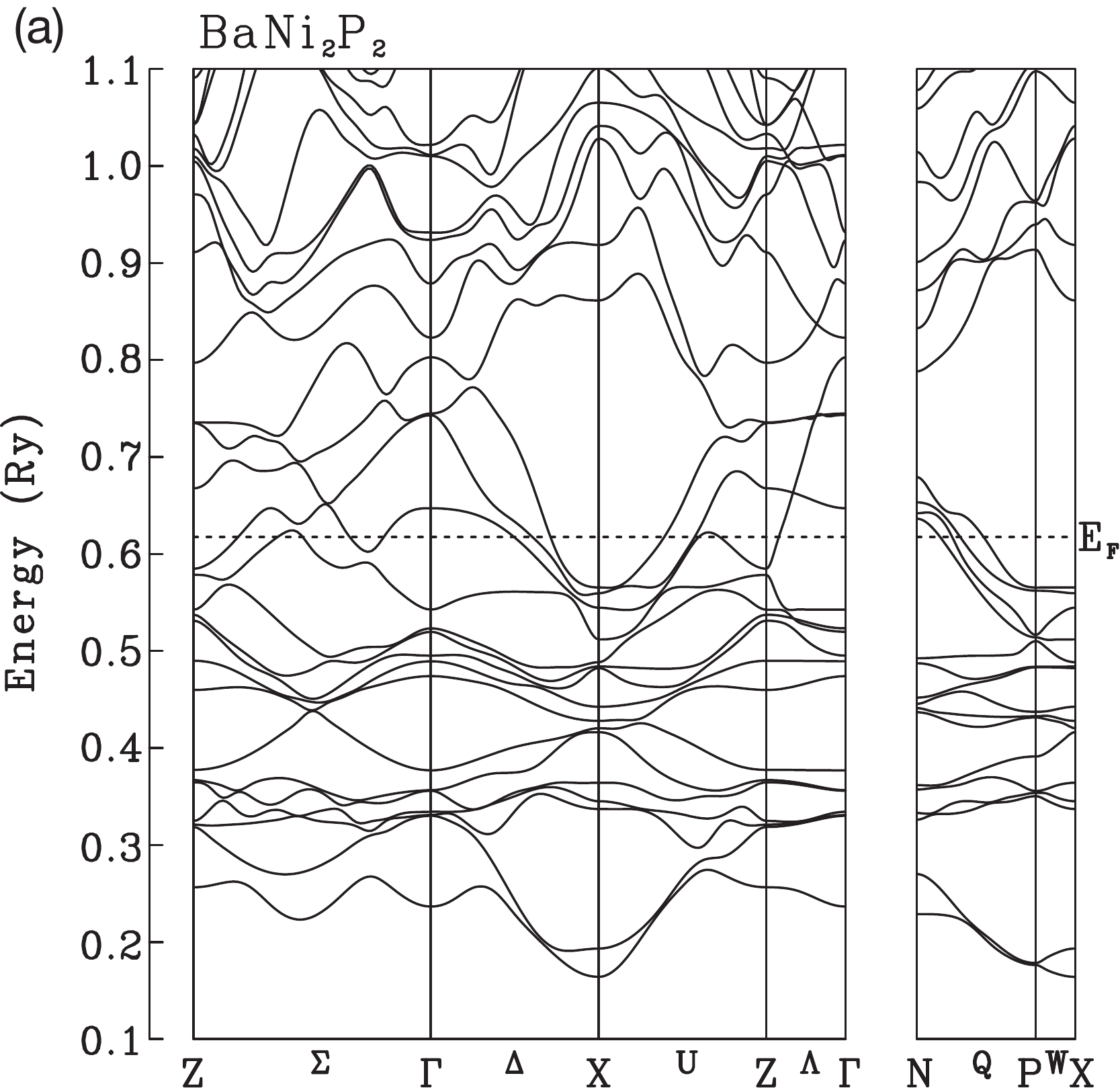}
\includegraphics[width=7cm]{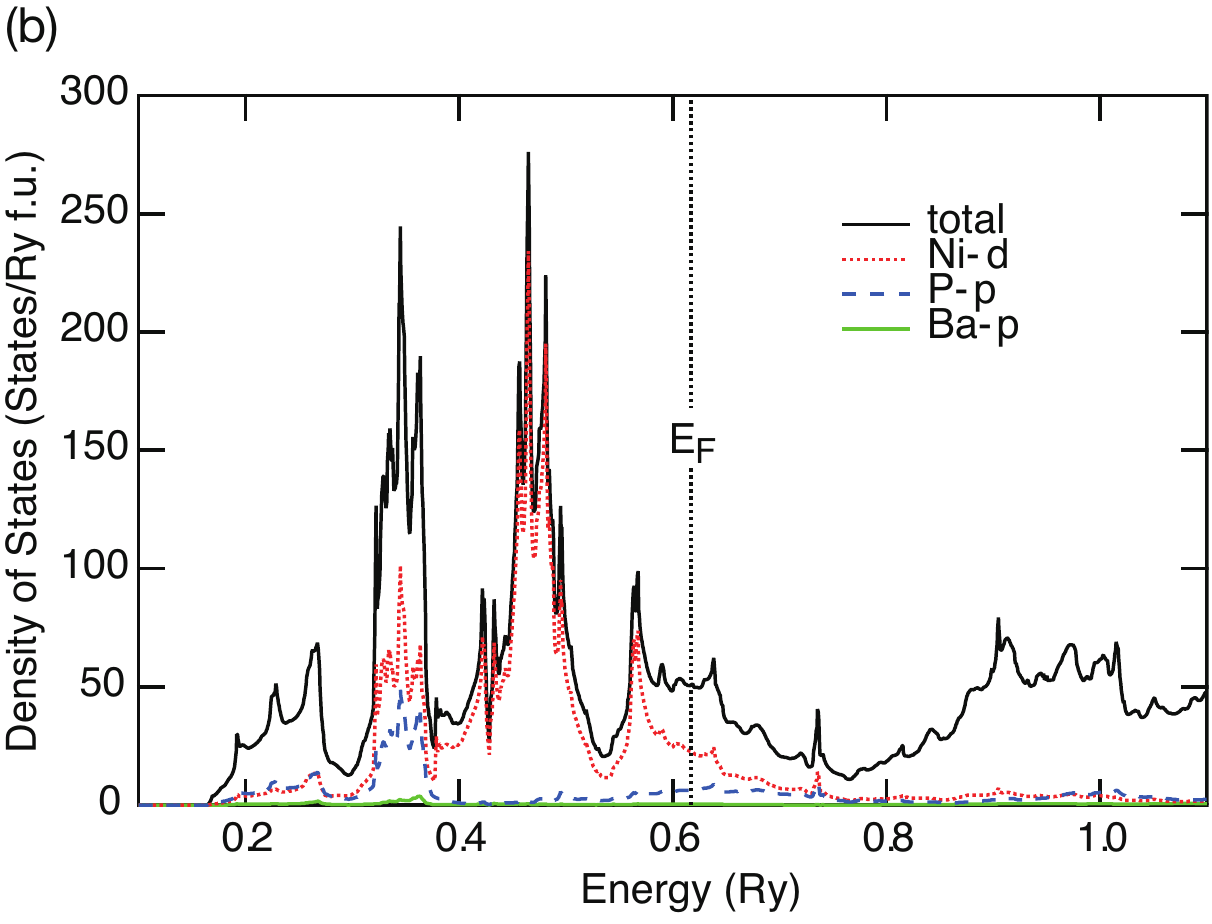}
\end{center}
\caption{(Color online) (a) Calculated electronic band structure along the high symmetry directions near the Fermi level $E_F$ of BaNi$_2$P$_2$.  (b) Total and partial densities of states.}
\label{f3} 
\end{figure}

Figures~\ref{f3}(a) and \ref{f3}(b) show the calculated electronic band structure and the total and partial densities of states near the Fermi level $E_F$ of BaNi$_2$P$_2$.  These results are in good agreement with previously published ones.\cite{Shameem99JAC}  Four bands 25--28 cross the Fermi level.  The density of states at $E_F$ is 50.77 states/Ry\,f.u., where f.u. is an abbreviation for formula unit.  This corresponds to the Sommerfeld coefficient of $\gamma = 8.78$ mJ\,/\,mol\,K$^2$, but no experimental value has been reported.  Approximately one-half of the density of states comes from the Ni-$d$ states.  The calculated Fermi surface is shown in Fig.~\ref{f2}(b).  The carrier numbers are 0.06 and 0.88 holes for band-25 and 26 hole surfaces, respectively, and 0.60 and 0.34 electrons for band-27 and 28 electron surfaces, respectively.  Note that BaNi$_2$P$_2$ is a compensated metal.  The theoretical dHvA frequencies calculated from this Fermi surface are shown in Fig.~\ref{f2}(a) and compared with the experimental ones.  Table~\ref{t1} also shows a comparison of the calculated frequencies and band masses $m_{band}$ with the experimental ones for the selected field directions.

Figure~\ref{f2}(a) indicates that all the four surfaces are experimentally observed and that the experimental and calculated dHvA frequencies are in excellent agreement: discernible deviations are found only for the smallest frequency branch $\alpha$.   Namely, it is experimentally confirmed that the LDA band-structure calculation can accurately describe the electronic structure near the Fermi level of BaNi$_2$P$_2$.  This contrasts sharply with the situation encountered when Fe-based compounds,  such as LaFeAsO, BaFe$_2$As$_2$, and LaFePO, are considered.  The electronic structures of these compounds are unusually sensitive to the atomic coordinate $z$ of As or P, and problems, such as the failure of structural optimization in nonmagnetic calculations and the overestimation of magnetic moment in magnetic calculations, are observed (see discussions in refs.~\citen{Mazin08PRB} and \citen{Ishibashi08JPSJS}).

As can be recognized from the band-26 hole surface [Fig.~\ref{f2}(b)], the Fermi surface of BaNi$_2$P$_2$ is three-dimensional.  This is another important contrast to the cases of the Fe-based compounds, in which the Fermi surface in the tetragonal structure is quasi-two-dimensional, composed of cylindrical surfaces, except for the possible existence of a small closed surface in LaFeAsO.\cite{Liu08PRL, Singh08PRL, Nekrasov08JETPLett}   This difference can basically be understood as follows.  Because of the hybridization between the transition-metal $d$ and the pnictogen $p$ states, the electronic bands near the Fermi level can be loosely grouped into bonding, nonbonding, and antibonding bands, corresponding to the three peaks of the density of states centered around 0.35, 0.47, and 0.6 Ry [Fig.~\ref{f3}(b)].  While the bonding and antibonding bands are three-dimensional owing to the hybridization, the nonbonding bands are essentially two-dimensional because of the minimal overlap between transition-metal $d$ orbitals along the $c$ axis.  The Fermi level of BaNi$_2$P$_2$ sits in the three-dimensional antibonding bands, whereas those of the Fe-based compounds are situated in the quasi-two-dimensional nonbonding bands.

The last column of Table~\ref{t1} shows that the dHvA mass enhancement $m^*/m_{band}$ is $\sim2$.  It is interesting to note that this value is similar to the enhancement of 2.3 found in the typical phonon-mediated strong-coupling superconductor Pb ($T_c$ = 7.2 K).\cite{Anderson72PRB}  In ref.~\citen{Subedi08PRB}, the electron-phonon coupling constant $\lambda_{ep}$ in BaNi$_2$As$_2$ has been evaluated from first-priciples calculations, and the corresponding mass enhancement $1+\lambda_{ep} = 1.76$ is close to our experimental value for BaNi$_2$P$_2$.  The authors of ref.~\citen{Subedi08PRB} argue that BaNi$_2$As$_2$ is a phonon-mediated superconductor and suggest that a role played by spin fluctuations, if any, is to suppress $T_c$ rather than to promote it.

Finally, we mention reports of dHvA measurements performed on related compounds.  For LaFePO, dHvA frequencies in the range $F\approx1\sim3$ kT have been observed.\cite{Sugawara08JPSJ, Coldea08PRL}  All the frequencies can be attributed to warped cylindrical Fermi surfaces and seem to basically be explained by band-structure calculations.  However, the quantitative agreement between the observed and calculated frequencies is rather poor, which is probably related to the difficulty in the band-structure calculations of the Fe-based compounds mentioned above.  The dHvA mass enhancement is estimated to be about four in ref.~\citen{Sugawara08JPSJ} and about two in ref.~\citen{Coldea08PRL}.  The factor-of-two difference again seems to suggest the difficulty in band-structure calculations.  It has also been reported that dHvA oscillations are observed in SrFe$_2$As$_2$.\cite{Sebastian08JPCM}  However, only small frequencies ($F\lesssim0.4$ kT) are reported, and since measurements are in the orthorhombic antiferromagnetic phase, a straightforward comparison with band-structure calculations or the present data is difficult.

In summary, we have performed dHvA measurements and a band-structure calculation for BaNi$_2$P$_2$.  All the four surfaces of the Fermi surface predicted by the calculation are experimentally observed, and the agreement between the experimental and calculated dHvA frequencies is excellent.  The determined Fermi surface is large and three-dimensional, in contrast to the Fermi surface in the Fe-based compounds.  The mass enhancement is about two, which is close to the theoretical value estimated for BaNi$_2$As$_2$.\cite{Subedi08PRB}

\section*{Acknowledgment}
The authors thank A. Sato (NIMS) for his help in determining the crystal axes of the samples.

\end{document}